\documentclass{emulateapj}

\usepackage{mathptmx}

\def\ergscm{erg~s$^{-1}$~cm$^{-2}$}

\def\arcdeg{\hbox{$^\circ$}}

\begin{document}

\submitted{Submitted to ApJ July 30, 2004; accepted February 9, 2005}

\title{Extragalactic source counts in the 20--50~keV energy band\\ from the
   deep observation of the Coma region by \emph{INTEGRAL}/IBIS}

\author{R.~Krivonos\altaffilmark{1}, A.~Vikhlinin\altaffilmark{1,2},
  E.~Churazov\altaffilmark{1,3}, A.~Lutovinov\altaffilmark{1},
  S.~Molkov\altaffilmark{1}, R.~Sunyaev\altaffilmark{1,3}}

\email{krivonos@hea.iki.rssi.ru}

\altaffiltext{1}{Space Research Institute, Moscow, Russia}
\altaffiltext{2}{Harvard-Smithsonian Center for Astrophysics, USA}
\altaffiltext{3}{Max-Planck-Institut f\"ur Astrophysik, Garching, Germany}

\shorttitle{EXTRAGALACTIC HARD X-RAY SOURCE COUNTS}
\shortauthors{KRIVONOS ET AL.}

\begin{abstract}
  We present the analysis of serendipitous sources in a deep, 500~ksec,
  hard X-ray observation of the Coma cluster region with the IBIS
  instrument onboard \emph{INTEGRAL}. In addition to the Coma cluster,
  the final 20--50 keV image contains 12 serendipitous sources with
  statistical significance $>4\,\sigma$. We use these data (after
  correcting for expected number of false detections) to extend the
  extragalactic source counts in the 20--50~keV energy band down to a
  limiting flux of $1.0\times 10^{-11}$~\ergscm{} ($\simeq
  1$~mCrab). This is a more than a factor of 10 improvement in
  sensitivity compared to the previous results in this energy band
  obtained with the \mbox{HEAO-1} A4 instrument. The derived source
  counts are consistent with the Euclidean relation, $N(>f)\propto
  f^{-3/2}$. A large fraction of identified serendipitous sources are
  low-redshift, $z<0.02$ AGNs, mostly of Seyfert 1 type. The surface
  density of hard X-ray sources is $(1.4\pm0.5)\times10^{-2}$ per square
  degree above a flux threshold of $10^{-11}$~\ergscm. These sources
  directly account for $\sim 3\%$ of the cosmic X-ray background in the
  20--50~keV energy band. Given the low redshift depth of our sample, we
  expect that similar sources at higher redshifts account for a
  significant fraction of the hard X-ray background. Our field covers
  only 3\% of the sky; a systematic analysis of other extragalactic
  \emph{INTEGRAL} observations can produce much larger source samples
  and is, therefore, critically important.
\end{abstract}

\keywords{X-rays: general --- X-rays: diffuse background --- galaxies: active --- galaxies: Seyfert}

\section{Introduction}

Most of the energy of the Cosmic X-ray Background (CXB) is emitted in
the energy band around 30~keV \citep{1980ApJ...235....4M}. However, the
exact nature of the source population responsible for the background at
these energies is unknown. The primary reason is low sensitivity of the
previous X-ray telescopes operating above 20 keV. Studies of the high
energy sources are also motivated by the recent work at lower energies,
2--10~keV. Most of the X-ray background at these energies is resolved
into sources \citep{2002ApJS..139..369G,2003AJ....126..539A} but the
spectrum of those sources does not match that of the CXB at high
energies, indicating the existence of a population of highly obscured
sources, which should be detectable more easily above 20~keV (e.g.,
Worsley et al.\ 2004 and references therein).

The telescopes onboard \emph{INTEGRAL} provide a major improvement in
sensitivity for X-ray imaging above 20~keV \citep{winkler2003}. During its
first year, \emph{INTEGRAL} has conducted a number of deep pointings to the
Galactic center region \citep{revnivtsev2004} and the Galactic plane
\citep{gps2003}, as well as to several extragalactic targets. One of the
deepest extragalactic observations was that of the Coma cluster region for a
total exposure of 500~ksec. The Coma cluster is located very close to the
North galactic pole and this field is minimally contaminated by the sources
within our Galaxy which dominate the hard X-ray sky. In addition, the image
is not ``polluted'' by a bright target and so it is excellent for detection
of faint sources.

In this Paper, we report on the analysis of faint serendipitous sources in
the Coma field. Our image reaches a $4\,\sigma$ detection threshold of
1~mCrab in the 20--50~keV energy band, above which we detected 12
serendipitous sources. Using this sample, we are able to extend the
extragalactic $\log N - \log S$ in the hard X-ray band to a flux limit of
$1.0\times 10^{-11}$~\ergscm, a factor of $\sim 10$ deeper than the
extragalactic part of the HEAO-1 A4 source catalog \citep{levine1984}.

\section{INTEGRAL observation of the Coma region}
\label{sec:obs}


The Coma region was observed by \emph{INTEGRAL} in 2003 on January 29--31
(revolution 36) for 170~ksec and on May 14--18 (revolutions 71--72) for
330~ksec. The observations consist of 221 shorter pointings which
form a $10^\circ \times 10^\circ$ grid around the target with a $2^\circ$
distance between the grid points. The January and May datasets have
different position angles, which leads to larger total sky coverage and
helps to minimize the systematic residuals in the final image. The data
quality in both series of observations is similar so they can be combined in
a single $\sim$500~ksec dataset.

We used the data from IBIS/ISGRI instrument which is the most suitable for
the imaging surveys in the hard X-ray band among the major instrument
onboard \emph{INTEGRAL}. IBIS \citep{ubertini2003} is coded-mask aperture
telescope. Its CdTe-based ISGRI detector \citep{lebrun2003} has a high
sensitivity above $20$~keV and has a high spatial resolution. The telescope
field of view is $28^{\circ} \times 28^{\circ}$ ($9^{\circ} \times
9^{\circ}$ full coded). All Coma pointings cover a $40^{\circ} \times
40^{\circ}$ region.

\begin{figure*}
  \centerline{
    \includegraphics[width=0.75\linewidth]{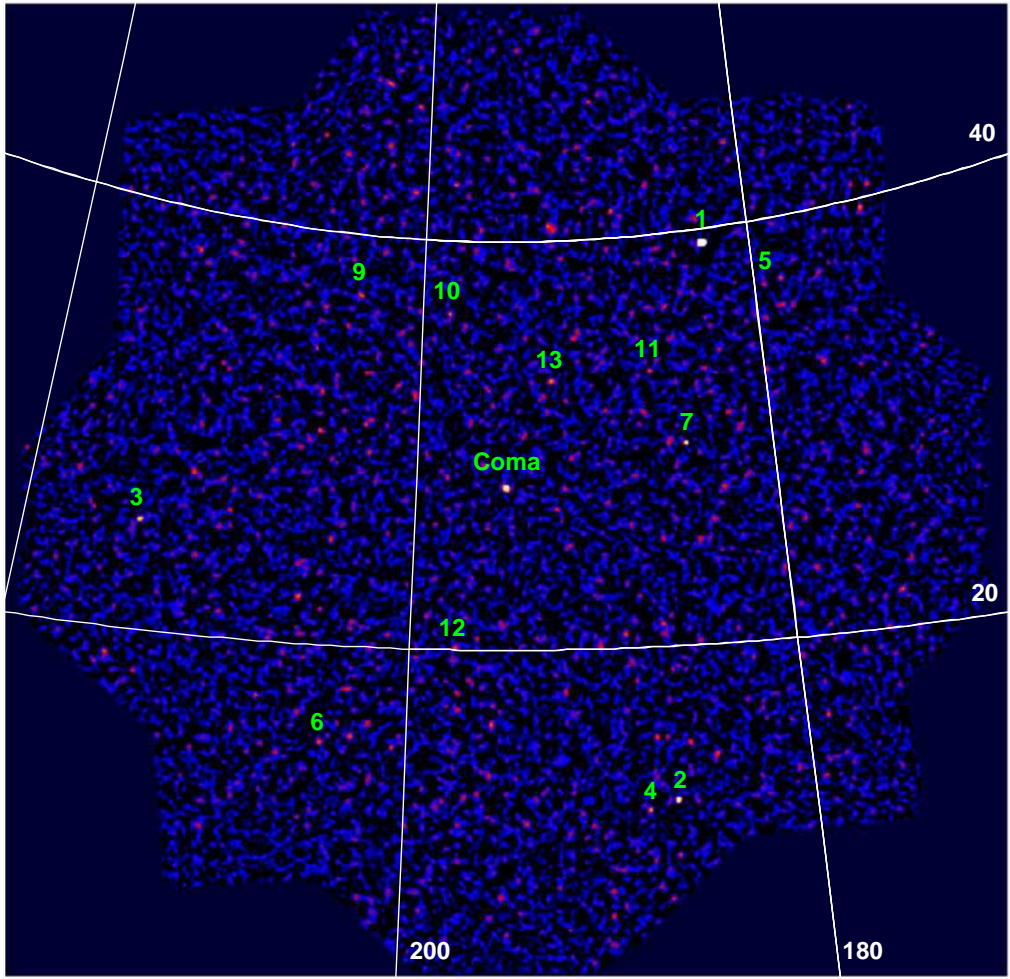}
    }
    \caption{
      X-ray image of Coma region in energy band 20-50~keV obtained by IBIS
      instrument onboard INTEGRAL during two different sets of
      observation. The color-coded intensity is in units of statistical
      significance for a point source. The markers indicate the locations of
      sources above a signal-to-noise level of 4.
    }
  \label{fig:img}
\end{figure*}

\section{Data reduction and image reconstruction}
\label{sec:data:reduction}

The IBIS data analysis involves special techniques for coded-mask image
reconstruction, for which we used the software suite developed by one of us
(EC). The essential steps are outlined below.

The event energies were calculated following the Off-line Scientific
Analysis software v3.0 (Goldwurm et al.\ 2003) using the gain table v9 and
the event rise time correction v7. Using the detector images accumulated in
a broad energy band, we searched for hot and dead pixels and then screened
the data to remove these artifacts. This resulted in rejection of several
percent of the detector area. We then computed raw detector images for each
pointing location and reconstructed the sky images individually.

The reconstruction starts with rebining the raw detector images into a
grid with the pixel size equal to 1/3 of the mask pixel size. This
is close but not exactly equal to the detector pixel size. Therefore,
the rebining causes a moderate loss of spatial resolution but leads to
straightforward application of the standard coded mask reconstruction
algorithms (e.g., Fenimore \& Cannon 1981, Skinner et
al. 1987a). Basically, the flux for each sky location is calculated as
the total flux in the detector pixels which ``see'' the location through
the mask, minus the flux in the detector pixels blocked by the mask:
\begin{eqnarray}
f=\sum_{M=1}D-BAL\sum_{M=0}D,
\label{eq:conv}
\end{eqnarray}     
where $f$ is the source flux, $D$ is the detector image, $M=1$ or $0$
corresponds to transparent or opaque mask pixels, respectively, and
$BAL\approx 1$ is so called balance matrix. The balance matrix accounts for
non-uniformity of the detector background. For a given location it is
calculated as:
\begin{equation}
  BAL=\frac{\Sigma_{M=1}D_b}{\Sigma_{M=0}D_b},
  \label{eq:bal}
\end{equation}  
where $D_b$ is the detector image accumulated over large number of
observations without strong sources in the field of view (i.e., the
background). Thus the expectation value of $f$ is zero for a source-free
field (see eq.~\ref{eq:conv},\ref{eq:bal}).

The images reconstruction is based on the DLD deconvolution procedure (see
notations in Fenimore \& Cannon 1981) when the mask pixel corresponds to
$n\times n$ detector pixels. The original detector is treated as $n\times n$
independent detectors and $n\times n$ independent sky images are
reconstructed and then combined into a single image. The point source in
such image is represented by a $n\times n$ square. In our case, this leads
to the effective Point Spread Function (PSF) being approximately a square of
$3\times3$ detector pixels or $12'\times12'$. 

%
  
Periodic structures in the mask produce sky images with a number of
prominent peaks accompanying every real source. An iterative removal
procedure was used to eliminate ``ghosts'' associated with the brightest
sources (NGC 4151, NGC 4388 and Coma). No iterative removal of sources was
performed for weak sources. The reconstructed images for each pointing grid
location were co-added in the sky coordinates. Finally, we computed the
statistical uncertainty of flux in the reconstructed image. This is a
straightforward procedure because the noise is Poisson in the original data
and it can be easily propagated through the image reconstruction algorithm
(i.e., data rebining and DLD deconvolution).

We used the 20--50~keV energy band for detection of the serendipitous
sources in the Coma field. The choice was motivated by the following
considerations. The lower boundary of the energy range where IBIS is
usefully sensitive is near 20~keV. The choice of the upper boundary is
driven by the desire to extend the energy band to the highest energy
possible without sacrificing the sensitivity. An upper boundary of 50~keV is
a reasonable choice because at higher energies, the IBIS sensitivity rapidly
decreases \citep{lebrun2003}. We checked that at higher energies, there are
no sources which are not detected in the 20--50~keV band. The reconstructed
image is shown in Fig.~\ref{fig:img}.

\section{Detection of Sources in the reconstructed image}

Most of the extragalactic sources are very faint in the hard X-ray band.
Their detection with \emph{INTEGRAL} is challenging and requires
sensitive techniques. Our approach is based on the so-called matched
filtering technique. A similar method applied to the source detection in
the \emph{ROSAT} PSPC images is extensively described in Vikhlinin et
al.\ (1995). We refer the reader to this paper and present only a brief
overview below.

\subsection {Matched Filter}

The matched filter source detection is based on the convolution of the
original image with the PSF. This method provides the most sensitive
detection algorithm for finding faint, isolated sources (e.g., Pratt
1978). The only difference between the traditional matched filtering and our
approach is to slightly modify the filter so that it can provide automatic
background subtraction.

The image reconstruction algorithm for coded-aperture imaging implies, in
principle, that the background in the reconstructed image is already
subtracted (see eq.~\ref{eq:conv}). However, some detector imperfections not
fully accounted for by the balance matrix and other practical problems might
lead to residual small-amplitude, large-scale intensity variations. The
automatic background subtraction is therefore a desirable property of the
detection algorithm.

The solution is to build the detection filter, $f(r)$, from two
components. The positive component in the center matches the PSF and
provides sensitive detection. The second one forms a negative ``ring'' at
larger radii so that the average value of $f(r)$ is zero. We decided to
implement $f(r)$ at the difference between the PSF and a Gaussian of larger
width,
\begin{equation}\label{eq:filter}
  f(r)=P(r)-A\exp\left(-\frac{r^2}{2a^2}\right), 
\end{equation}
where $P(r)$ is the IBIS PSF, and the amplitude satisfies
\begin{equation}\label{eq:filter:gauss}
  A=\frac{1}{a^2} \int_{0}^{\infty} P(r)\,r\,dr.
\end{equation}

\begin{figure}
  \centerline{ \includegraphics[width=0.95\linewidth]{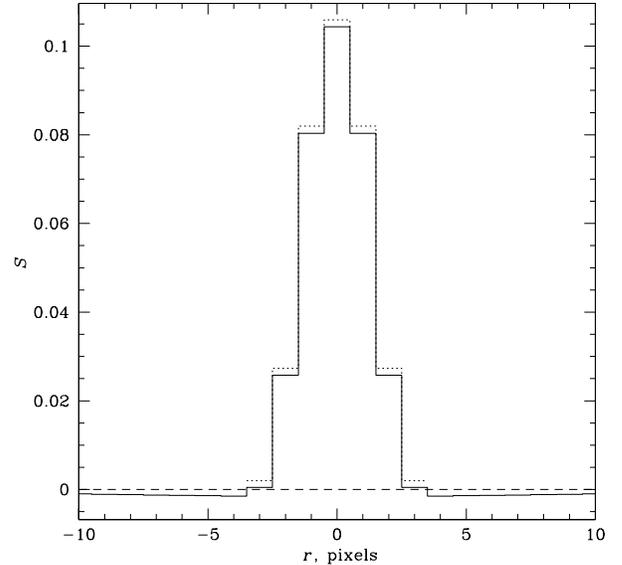} }
  \caption{One-dimensional slice through the PSF of our reconstructed
  images after iterative source removal procedure (dotted histogram)
  and the profile of the detection filter (solid histogram).  The
  pixel size is $4'$.  The PSF is the result of coadding of many real
  images (Crab Nebula) with different orientations. The resulting
  effective PSF is then more round and smooth than the ideal PSF
  (single pointing, no rebinning, perfect detector) which is a flat 3
  by 3 pixels square in the DLD sky. } \label{fig:psf}
\end{figure}

The PSF calibration was obtained from the analysis of Crab
observations. Crab is essentially a point source for \emph{INTEGRAL} (the
source size smaller than $4'$ or 1 detector pixel), therefore its image can
be used as a representation of the PSF (we actually used the average of 4
possible reflections of the Crab image around the $X$- and $Y$-axis to
reduce statistical uncertainties). The derived PSF is essentially a square
of $3\times3$~pixels (or $12'\times12'$) surrounded by fainter larger-scale
wings; its one-dimensional slice is shown in Fig.~\ref{fig:psf}.

Given the PSF, the only free parameter in the filter defined by
eq.~(\ref{eq:filter},\ref{eq:filter:gauss}) is the Gaussian width, $a$. The
choice of $a$ is driven by two conflicting requirements: a) the larger
values of $a$ lead to better signal-to-noise (SNR) ratios in the convolved
images, and b) smaller values of $a$ lead to more accurate subtraction of
small-scale background variations. We chose $a=40'$ which results in the
signal-to-noise ratio of 99\% of its maximum value (that for
$\sigma\rightarrow\infty$).

\subsection{Source Detection, Fluxes, Locations}
\label{sec:det:flux:loc}

The sources are identified as local maxima in the the filter-convolved
image. To sort out statistical fluctuations, we computed the convolved
noise as
\begin{equation}\label{eq:noise:map}
  N(x,y) = \left((\text{noise map})^2\otimes(f^2)\right)^{1/2}.
\end{equation}
This equation is strictly correct only if the individual image pixels
are statistically independent, which is not fulfilled in our case
because of rebinning of the detector images
(\S~\ref{sec:data:reduction}). However, the effect is small because the
correlation length of the noise in the reconstructed images is smaller
than the size of the filter. It can be corrected by uniform rescaling of
the noise map given by eq.~(\ref{eq:noise:map}) so that the \emph{rms}
variation of the ratio of the convolved image and $N(x,y)$ is 1. The
required correction factors are $\approx 0.9$. Note that such adjustment
effectively corresponds to noise determination directly from the data,
which is the most trustworthy approach.

Accepted sources are the maxima in the convolved image with the
signal-to-noise ratio $>~4$. This threshold is lower than what is often
used in wide-area surveys. Our choice was motivated by the following
considerations. Detection threshold is the result of compromise between
our desire to exclude as many false detections as possible (this drives
the threshold high) while retaining the maximum of real sources in the
sample (this drives it low). The $4\sigma$ detection threshold results
in manageable contamination of the sample by false detections, $4.7$ on
average or $\sim 33\%$ of the total sample, which still can be reliably
subtracted in the final $\log N - \log S$ analysis
(\S~\ref{sec:false:detections}).

Source fluxes can be obtained directly from filter-convolved images. The
peak value in the convolved image is proportional to total source flux
because convolution with the detection filter is a linear image
transformation.  The conversion factor between the peak value and flux
was obtained using the reconstructed Crab images, for which we assumed
the Crab spectrum $9.7\times
E_{\text{keV}}^{-2.1}$~phot~cm$^{-2}$~s$^{-1}$~keV$^{-1}$
\citep{1974AJ.....79..995T}. The conversion factor is a function of
position because of several effects. First, the PSF degradation at large
off-axis angles changes the peak value in the filter-convolved images.
Second, off-axis PSF contains weak side-lobes which leads to
over-subtraction of the source flux by a wavelet-like filter with a
negative annulus, such as ours. Third, there is s vignetting effect most
likely caused by absorption in the mask holes. These effects were
calibrated using a large number of Crab observations with the source at
different locations within the FOV. These observations were reduced
identically to the Coma field. The derived peak-to-flux coefficient
decreases by $22\%$ from $0^{\circ}$ to $5^{\circ}$ off-axis and then
stays approximately constant. We used the conversion factor derived at
$5^{\circ}$ and ignored the trend at smaller off-axis angles because it
affects only a small fraction of the image area.

The source locations were measured as flux-weighted mean coordinates
within $6'$ of the maxima in the filter-convolved images. The accuracy
of this method was tested using a large number of INTEGRAL
observations in which there were sources with $\mathrm{SNR}=5-6$. The
distribution of coordinate offsets for these sources is well described
by a Gaussian with a 68\% uncertainty radius of $4.2'$.

The noise in the reconstructed image increases near the edge of the
field of view. This region, if included in the survey, can pollute the
source catalog with high-flux false sources. Therefore, we excluded the
regions where the \emph{rms} level of the image noise is a factor of 10
greater than that in the center of the field of view. This reduced the
image area by 28\%.

\section{Detected Sources}

\begin{deluxetable*}{p{1.7cm}cccrcccccc}
\tablewidth{0pt}
\tablecaption{Hard X-ray sources detected in the Coma field
  \label{tab:lst}
  \label{tab:opt:id}
}
\tablehead{\colhead{X-ray ID}&
\colhead{RA} & 
\colhead{Dec} & 
\colhead{Flux} & 
\colhead{SNR} &
\colhead{Optical ID} &
\colhead{Dist.} &
\colhead{Type} &
\colhead{$z$} &
\colhead{$L_x$} &
\colhead{\emph{ROSAT}}\nl
\colhead{} & 
(J2000) & 
(J2000) & 
\ergscm &
\colhead{} &
\colhead{} &
\colhead{arcmin} &
\colhead{} &
\colhead{} &
\colhead{erg~s$^{-1}$}
}
\startdata
1\dotfill &12 10 33&+39 24 21& $(3.2\pm0.1)\times10^{-10}$ &44.5& NGC 4151 &
0.1 &Sy1.5  & 0.0033&$7.3\times10^{42}$&+\\
2\dotfill &12 25 43&+12 40 42& $(1.1\pm0.1)\times10^{-10}$ &9.5&NGC 4388 &
1.3 & Sy2 & 0.0084&$1.6\times10^{43}$& \\ 
3\dotfill &14 17 51&+25 08 43& $(5.3\pm1.2)\times10^{-11}$ & 5.1& NGC 5548 &
2.0& Sy1    & 0.0171&$3.3\times10^{43}$&+\\ 
4\dotfill &12 31 19&+12 14 01& $(5.2\pm1.3)\times10^{-11}$ & 4.7& \nodata  & \nodata&\nodata&\nodata& \\ 
5\dotfill &11 56 47&+37 00 20& $(3.3\pm1.0)\times10^{-11}$ & 4.0& \nodata  & \nodata&\nodata&\nodata& \\ 
6\dotfill &13 37 09&+15 20 00& $(2.9\pm0.8)\times10^{-11}$ & 4.1& \nodata  & \nodata&\nodata&\nodata& \\ 
7\dotfill &12 18 23&+29 49 06& $(1.7\pm0.3)\times10^{-11}$ & 5.8& NGC 4253
& 0.8 &  Sy1    & 0.0129&$5.9\times10^{42}$&+\\ 
8\dotfill &12 59 35&+27 57 25& $(1.6\pm0.2)\times10^{-11}$ &
7.7&Coma\tablenotemark{1}& 3.3 & GClstr &  0.0244&$2.4\times10^{43}$&+\\ 
9\dotfill &13 35 12&+37 12 17& $(1.5\pm0.4)\times10^{-11}$ & 4.1& V$^*$ BH
CVn & 3.0 & CV     &\nodata&\nodata&+ \\ 
10\dotfill&13 13 23&+36 34 20& $(1.6\pm0.3)\times10^{-11}$ & 4.6& NGC 5033 &
1.6 & Sy1    & 0.0029&$2.4\times10^{41}$&+\\ 
11\dotfill&12 25 40&+33 31 13& $(1.4\pm0.3)\times10^{-11}$ & 4.2& NGC 4395 &
2.5 & Sy1    & 0.0011&$3.0\times10^{40}$&+\\ 
12\dotfill&13 10 18&+20 01 51& $(1.1\pm0.3)\times10^{-11}$ & 4.2& \nodata  & \nodata&\nodata&\nodata& \\ 
13\dotfill&12 48 49&+33 15 14& $(1.0\pm0.3)\times10^{-11}$ & 4.3& \nodata  & \nodata&\nodata&\nodata&     
\enddata

\tablecomments{The uncertainty in the source locations is
  approximately $4'$ (90\% confidence). The fluxes and luminosities
  are in the 20--50~keV energy band.}

\tablenotetext{1}{The counts-to-flux conversion for Coma is inaccurate
  because the source is extended and has the thermal spectrum.
  \emph{INTEGRAL} data on Coma will be discussed elsewhere.}

\end{deluxetable*}

Thirteen hard X-ray sources were detected above the $4\sigma$ threshold
in our Coma image, including the target. The source locations and
observed fluxes in the 20--50~keV energy band are given in
Table~\ref{tab:lst}. We searched for obvious optical identifications of
our sources using the NASA Extragalactic Database as well as by the
visual inspection of the images from Digitized Sky Survey and the
\emph{ROSAT} All-Sky Survey. Seven sources were unambiguously identified
with the extragalactic objects (Table~\ref{tab:opt:id}), most of them
classified as Seyfert1 galaxies at low redshifts, $z<0.02$.

Our brightest sources were observed in hard X-rays with previous
observatories \citep{1999ApJS..120..335M}. NGC 4151 was observed with
HEAO-1 A4 \citep{1984ApJ...279..555B}, SIGMA
\citep{1994ApJS...92..343M,1995A&A...300..101F}, BeppoSAX
\citep{1998axrs.symp..481P}, BATSE \citep{1998ApJ...501..608P}.
Seyfert-1 galaxy NGC 5548 was detected by HEAO-1 A4
\citep{1983ApJ...269..423R}, BeppoSAX \citep{2000ApJ...536..718N}, BATSE
\citep{2000ApJS..127...79L}. NGC 4388 was detected by SIGMA
\citep{1992A&A...264...22L} and BATSE \citep{1996A&AS..120C.559B}.

Most of positively identified \emph{INTEGRAL} sources are also detected
in the soft X-ray band in the \emph{ROSAT} All-Sky Survey. The only
exception is the Sy2 galaxy NGC~4388. This second-brightest
\emph{INTEGRAL} source is undetectable in the \emph{ROSAT} energy band,
most likely because of a high intrinsic absorption
\citep{2004ApJ...614..641B}. One source (\#9) is identified with the
X-ray bright cataclysmic variable in our Galaxy. No identified sources
have the same redshift as the Coma cluster, the observation
target. Therefore, they are not located within the cluster or associated
large-scale structures and we can safely use the entire sample to derive
the serendipitous source counts.

We were unable to unambiguously identify 6 detected sources, partly because
of the relatively poor positional accuracy achievable in the hard X-ray
band. We note that some fraction of the unidentified sources are likely to
be false detections as discussed below.

\section{log N -- log S for detected sources}

\subsection{Survey Area}

\begin{figure}[tb]
  \centerline{
    \includegraphics[width=0.95\linewidth]{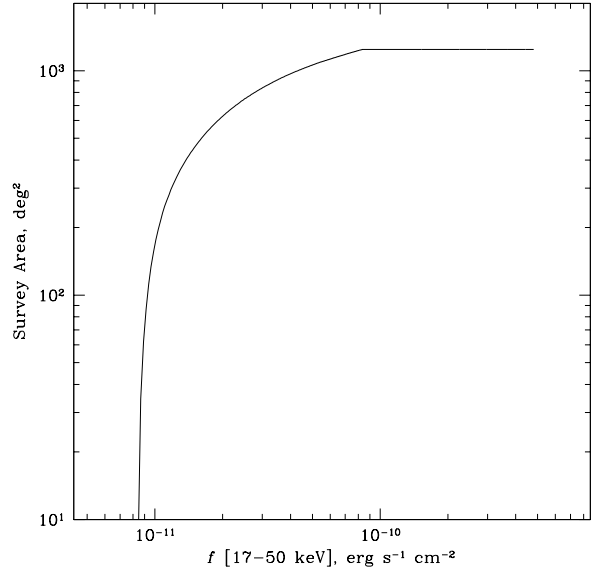}
    }
  \caption{Survey area as a function of flux for sources with $\mathrm{SNR}>4$.}
  \label{fig:area}
\end{figure}

Since the image noise systematically increases at large off-axis distances,
our detection threshold of $\mathrm{SNR}=4$ corresponds to different fluxes
at different locations. Therefore, the survey area is a function of flux. 
This function is computed easily for the given noise map because the peak
values in the convolved image is simply proportional to the source flux (see
\S\ref{sec:det:flux:loc}). 

Equation~(\ref{eq:noise:map}) was used to obtain the noise map for the
filter-convolved image. Multiplying this map by 4 and then by the flux
conversion coefficient gives us the map of the minimal detectable flux,
$f_{\text{min}}$, as a function of position. By counting the area where
$f>f_{\text{min}}$, we obtain the survey area as a function of flux,
$A(f)$. The results of this computation are shown in
Fig.~\ref{fig:area}.  The minimum detectable flux at the very center of
the field of view is $0.9\times10^{-11}$~\ergscm{} (or $\simeq
0.9$~mCrab). The survey area reaches its geometric limit of 1243 square
degrees for $f>8.3 \times10^{-11}$~\ergscm, and 50\% of this area has
the sensitivity better than $f=1.9 \times 10^{-11}$~\ergscm.

Given the survey area as a function of flux, $A(f)$, the cumulative source
counts (also referred to as the $\log N - \log S$ distribution) can be
computed easily as
\begin{equation}
  \label{eq:logn-logs}
  N(>f) = \sum_{f_i>f} A(f_i)^{-1}.
\end{equation}
However, our source catalog must contain a small number of false detections
because of the relatively low detection threshold ($\mathrm{SNR}=4$). We
need to subtract their contribution from the $\log N - \log S$ measurement.

\subsection{Expected Number of False Detections}
\label{sec:false:detections}

The reconstruction algorithm for the coded-aperture images leads to the
symmetric, Gaussian distribution noise in the resulting image. Therefore,
the number of false detections (those arising because of noise) above a
signal-to-noise ratio $\sigma$ can be estimated by counting the local minima
with the amplitude below $-\sigma$.

\begin{figure}[tb]
  \centerline{
    \includegraphics[width=0.95\linewidth]{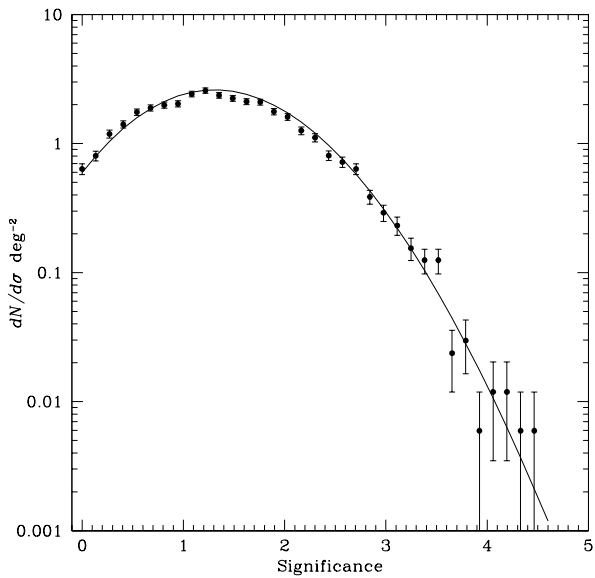}
    }
  \caption{
    Distribution of the formal statistical significance of the local minima
    in the filter-convolved \emph{INTEGRAL} image. The solid line shows the
    fit derived from Monte-Carlo simulations of  the images with Gaussian
    noise, slightly rescaled to match the observed distribution.
  }
  \label{fig:min_fit}
\end{figure}

The small number of the local minima with sufficiently high formal
significance ($\mathrm{SNR}\lesssim-4$) is the main practical
difficulty. We, therefore, adapted the following approach. We simulated a
large number of images with the Gaussian noise, convolved them with the
detection filter, and derived the distribution of the local maxima as a
function of their formal significance, $\sigma$. The distribution was fit to
a third-order polynomial in the $(\sigma,\log(dN/d\sigma))$ coordinates. The
analytic model derived from simulations was slightly rescaled to match the
distribution of local minima in the real data (Fig.~\ref{fig:min_fit}).

The rescaled model allows us to predict the number of false detections as a
function of the statistical significance threshold. We can also use it to
predict the number of false detections as a function of limiting flux
because we know the image noise as a function of position. Above our
limiting flux, $1.0\times10^{-11}$~\ergscm, we expect 4.7 false detections,
and 1.7 false sources above $2.8\times10^{-11}$~\ergscm. The expected
average number of false detections is less than 0.1 at fluxes
$f>10^{-10}$~\ergscm. We can also compute the corresponding $\log N - \log
S$ function for false sources using the survey sky coverage, identically to
what is done with the real sources,
\begin{equation}\label{eq:logn-logs-false}
  N_{\text{false}}(>f) =
  \sum_{i,j}\int_{\sigma_{\text{min}}}^{\infty}\frac{dN/d\sigma}{A(\sigma\,
  n_{i,j})} \,d\sigma,
\end{equation}
where $n_{i,j}$ is the image noise in the pixel $(i,j)$,
$\sigma_{\text{min}} = \mathrm{max}(4,f/n_{i,j})$, $A(f)$ is the sky
coverage as a function of flux, $dN/d\sigma$ is the number of false
detections in one pixel as a function of the formal statistical
significance, and the sum is over all image pixels.

This procedure for estimating the number of false detections has been
checked using shorter, 64~ksec, pieces of the Coma observation. Those
sources detected in the short-exposure image and absent in the total
500~ksec image, are false. We verified that the $\log N - \log S$ for false
sources predicted by eq.~(\ref{eq:logn-logs-false}) for short exposures is
in excellent agreement with that observed.

%
%
%

\subsection{Results}

\begin{figure}[tb]
  \centerline{
    \includegraphics[width=0.95\linewidth]{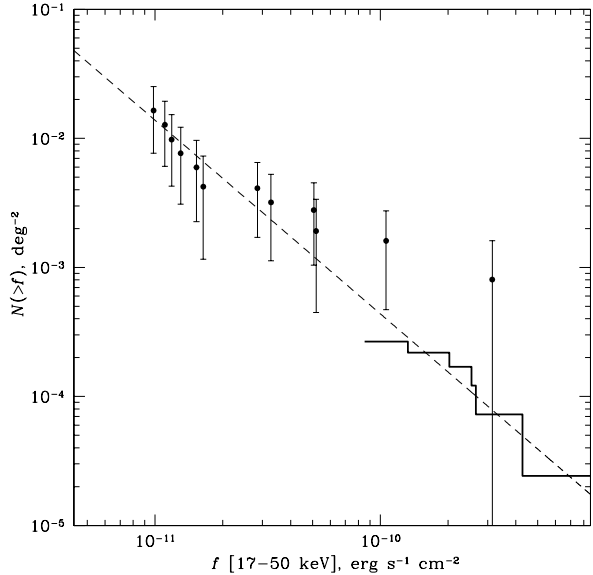}
  }
  \caption{The $\log N -  \log S$ distribution for the serendipitous
    hard X-ray sources in the Coma field. The histogram shows the source
    counts derived from the HEAO-1~A4 extragalactic sample
    ($|b|>30^{\circ}$). The HEAO-1~A4 fluxes were converted to our energy
    band from the nominal 25--40~keV band assuming a power-law spectrum with
    $\Gamma=2$. The dashed lines shows a $N\propto f^{-3/2}$ fit to our
    dataset.}
  \label{fig:logn-logs}
\end{figure}

Figure \ref{fig:logn-logs} shows the $\log N - \log S$ distribution
for the serendipitous sources in the Coma field, corrected for the
survey sky coverage as a function of flux, and subtracted contribution
of false detections. Our results can be compared directly only with
the previous measurements with the \mbox{HEAO-1 A4} instrument which
operated in the overlapping energy band, 25--40~keV
\citep{levine1984}.

The $\log N - \log S$ for the HEAO-1 A4 source catalog in the
extragalactic sky ($|b|>30^{\circ}$) is shown by the solid histogram
in Fig.~\ref{fig:logn-logs}. A single Euclidean function,
$N(>f)\propto f^{-3/2}$, can fit both datasets. This is not surprising
given that our sources are at low redshifts. For the fixed slope of
the $\log N - \log S$ distribution at $-3/2$, we derive from our data
only the surface density of the extragalactic hard X-ray sources
$N=(1.4\pm0.5)\times10^{-2}$ per square degree above a limiting flux
of $10^{-11}$~\ergscm~in the 20--50~keV energy band.

All identified INTEGRAL sources have low redshifts, $z<0.02$. Therefore,
the depth of our survey is small and it can be affected by the nearby
large-scale structures. It would be extremely important to extend such
measurements to a larger number of fields. However, it is interesting to
note that, thanks to a greater sensitivity of INTEGRAL, the volume
covered by our observation is larger than the volume covered by the
extragalactic ($|b|>30\deg$) portion of the HEAO-1 A4 all-sky survey.

The all-sky catalog from the \emph{RXTE} slew survey
\citep{2004A&A...418..927R} reaches a similar depth as our survey,
although at a lower energy band of 8--20~keV. We can compare the number
densities of sources by converting the \emph{RXTE} fluxes to our energy
band assuming a power law spectrum with $\Gamma=1.6$, the typical photon
index for the \emph{RXTE} AGNs (Fig.~8 in Revnivtsev et al.). The
\emph{RXTE} $\log N - \log S$ predicts the source density of
$(0.9\pm0.1)\times10^{-2}$ per square degree above the flux limit of our
survey; this is somewhat lower than, but within the uncertainties of,
our measurement.

\section{Discussion and conclusions}

We have presented the analysis of the deepest hard X-ray image of the
extragalactic sky obtained to date. We detected 12 serendipitous sources
in the $40^\circ\times40^\circ$ field centered at the Coma cluster. Most
of the detected sources appear to be the Sy1 galaxies at low redshifts,
$z<0.02$.

The $\log N - \log S$ distribution derived from a combination of our
survey and extragalactic sample of the \mbox{HEAO-1 A4} source catalog
follows the Euclidean function, $N(>f)\propto f^{-3/2}$. The observed
normalization of the $\log N - \log S$, $(1.4\pm0.5)\times10^{-2}$~per
square degree above a limiting flux of $10^{-11}$~\ergscm, corresponds
to the integrated flux of $4.3\times10^{-13}$~\ergscm~deg$^{-2}$, or 3\%
of the total intensity of the hard X-ray background in the 20--50~keV
energy band \citep{1980ApJ...235....4M}.

The redshift depth of our source catalog appears to be below $z=0.02$.
Therefore, it is reasonable to expect that the Euclidean source counts
will extend to much fainter fluxes than our sensitivity limit. For
example, the depth of $z<0.2$ where the space curvature of AGN evolution
effects are still small, will correspond to approximately a factor of
100 lower flux. Extrapolation of our $\log N - \log S$ by a factor of
100 towards fainter fluxes will account for 30\% of the total X-ray
background in the 20--50~keV energy band. In short, we start to uncover
the source population responsible for a significant fraction, if not
most, of the hard X-ray background.

\acknowledgments

We are pleased to acknowledge the efforts of the Russian \emph{INTEGRAL}
Science Data Center to facilitate the data distribution and scientific
analysis of our observation. We thank M.~Revnivtsev and S.~Sazonov for
careful reading of the manuscript and comments. Financial support was
provided by the grant NSH-2083.2003.2 from the Russian Ministry of
Science, by the Department of Physics of the Russian Academy of
Sciences, by Russian Basic Research Foundation (grants 02-02-16619), and
by \emph{NASA} grant NAG5-9217. RK was also supported by RBRF grant
03-02-17286.

\end{document}